\documentclass{appolb}
\usepackage{epsfig,wrapfig,psfrag}

\newcommand{\be}{\begin{equation}}
\newcommand{\ee}{\end{equation}}
\newcommand{\ba}{\begin{eqnarray}}
\newcommand{\ea}{\end{eqnarray}}

\newcommand{\bi}{\bibitem}
\begin{document}

\title{Spin-dependent, interference and $T-$odd fragmentation and fracture
functions
}
\author{O.~V.~Teryaev,\address{
 Joint Institute for Nuclear Research, Dubna, 141980 Russia}}
\maketitle
\begin{abstract}
Fracture functions, originally suggested to describe
the production of diffractive and leading hadrons in
semi-inclusive DIS, may be also applied at fixed target energies.
They may also include interference and final state interaction,
providing a source for azimuthal asymmetries at HERMES and
(especially) $\Lambda$ polarization at NOMAD.
The recent papers by Brodsky, Hwang and Schmidt, and by
Gluck and Reya, may be understood in terms of fracture functions.
\end{abstract}
\section{Introduction}
QCD factorization allows one to express the cross-sections and
polarization observables  of hard processes in terms of
convolutions of partonic subprocess and non-perturbative
functions, describing the hadron-parton and parton-hadron
transitions.
The studies of various spin effects results in the extension of
their possible types. As usual, the case of Single Spin Asymmetries (SSA)
is especially difficult, requiring the
interference and final state interactions, producing the imaginary phase.
The most widely known objects are parton distributions, describing
the fragmentation of hadrons to partons and related to the
forward matrix elements
$\sum_X<P|A(0)|X><X|A(x)|P>=<P|A(0)A(x)|P>$
of renormalized non-local light-cone quark and gluon operators.
As they do not contain any variable, providing the cut and
corresponding imaginary phase (to put it in the dramatic manner,
the proton is stable), the T-odd distribution functions can not
appear in the framework of the standard factorization scheme. At
the same time, they may appear effectively, when the imaginary
phase is provided by the cut from the hard process, but may be
formally attributed to the distribution \cite{BMT}.
Another well-known object is fragmentation function, describing
the fragmentation of partons to hadrons and constructed from the
time-like cutvertices of the similar operators
$\sum_X <0|A(0)|P,X><P,X|A(x)|0>$.  Now, they may contain the cut with
respect to the time-like parton momentum squared $k^2$(which was
space-like in the case of distributions), corresponding, at the
hadronic level, to the jet mass. This may give rise to the
number of T-odd fragmentation functions, including jet
handedness \cite{hand}, Collins function \cite{col} and
interference fragmentation functions \cite{jt}.

The FRACTURE function (FF)\cite{Tren}, whose particular example
is represented by the diffractive distribution (DD)\cite{coll2},
is related to the object
$\sum_X <P_1|A(0)|P_2,X><P_2,X|A(x)|P_1>,$
combining the properties of FRAgmantation and struCTURE
functions. They describe the correlated fragmentation of hadrons
to partons and vice versa. Originally this term was applied to
describe the quantities integrated over the variable
$t=(P_1-P_2)^2$, while the fixed $t$ case is described by the
so-called extended fracture functions \cite{Gra}.

\section{Interference and T-odd fracture functions}
They may be also extended \cite{TODD}
to describe SSA in such processes.
Namely, such functions can easily get the imaginary phase from the cut
produced by the variable $(P_1+k)^2$. Due to the extra momentum
of produced hadron $P_2$, the number of the possible P-odd
combinations increases. Therefore, they may naturally allow for
the T-odd counterparts.

The T-odd part of
(inclusive) DIS was studied long ago, when the non-local
analysis of twist 3 terms was presented for the first time
\cite{ET84}. As soon as DIS does not contain any cuts, these
effects require the real T-violation and are of a pure academic
interest for the foreseen future of spin experiments.
At the same time, the similar effects for the crossing related
process of semi-inclusive annihilation correspond to the
distributions substituted by fragmentation functions. As the
latter may contain the imaginary cuts, simulating the
T-violation, the performed calculation is starting to be more
related with physics. Namely, it describes the production of
transverse polarized baryon (one should typically think about
$\Lambda$, whose polarization is easily revealed in its weak
decay) in the annihilation of the unpolarized leptons
\cite{spin96}.
The consideration of TOFF is actually completely similar.
One should just substitute the transverse polarization of the
baryon by the product of the transverse component of produced
particle momentum and the longitudinal polarization of the
initial particle $s_T \to P_{2 T} s_L/M $. Such a simultaneous
appearance of the momentum and polarization of the different
particles is the natural consequence of the correlated
fragmentation of hadrons to partons and vice versa, described by
FF. The resulting expression for the hadronic tensor, combining
the contributions of quark and quark-gluon TOFF's is the
straightforward counterpart of that for annihilation of
unpolarized leptons (see (18) of \cite{spin96}), up to the
mentioned substitution and the change of fragmentation function
$c_V$ to the TOFF $F(x,\xi,t)$. The longitudinal proton
polarization ($S_L$)-dependent part is taking the following
form:
\begin{equation}
\label{dis}
W^{\mu \nu}=\frac{s_L}{Q^2} \sum_{i=q,\bar q} e^2_i x_B
F_i(x_B,\xi,t)[(2x_B P_1^\mu+q^\mu) \epsilon^{\nu P_1 P_2 q}+
(2x_B P_1^\nu+q^\nu) \epsilon^{\mu P_1 P_2 q}].
\end{equation}
The case of polarized partons, rather than hadrons,
corresponds to the matrix elements of axial, rather than vector
operators. Another generalization may be provided by the case of
the multihadron fragmentation. It is this latter case,
considered by J. Collins as a "polarized beam jets"
\cite{coltod}, which is the first description of TOFF.
In the case of the produced baryons, rather than pions, the number of
possible TOFF substantially increases. In the case of
unpolarized target, the direction of
transverse polarization of produced $\Lambda$
may be defined by both lepton and hadron scattering planes.
It is the former case, which may be described by the
same expression (\ref{dis}), with pion momentum substituted by $\Lambda$
transverse polarization, which results in return to the mentioned
formula of \cite{spin96}.
Note that full set of T-odd fracture functions may be
studied along the line discussed here \cite{blum} by
dropping the requirement of T-invariance (as T violation may be
{\it simulated} by imaginary phases from FSI).

Let us now discuss the possible experimental manifestations
of these effects.

\section{T-odd Fracture Functions at
HERMES and NOMAD}

First point, which should be mentioned in this connection, is
the necessity for minor generalization of FF. Namely, one should
consider the possibility of the hadron 2 being different from
the hadron 1 (pion for HERMES and $\Lambda$ for NOMAD).
This generalization is in fact straightforward and
do not require any changes in the proof of factorization.

One may also worry, why the correlated fragmentation could be
important for the hadrons, which are produced in the current,
rather than target, fragmentation region, studied by HERMES.
This generalization is more serious.  It is based on the fact,
that the invariant measure of such a correlation is provided by
the squared momentum transfer $t=-Q^2z/x$, which can be rather
small for HERMES and NOMAD kinematics. Of course part of that smallness
comes from the smallness of $Q^2$, but it is well known, that
because of "handbag dominance" the scaling in $Q^2$ happens much
earlier than in $t$. Consequently, the corrections to the
factorized distribution and fragmentation functions, provided by
fracture functions, may be important, especially at lower $z$.

One should mention in this connection the successful
application of handbag dominance in the area of GPD, having, as
it was mentioned above, much in common with FF. Namely, it is a
description of large-angle (real) Compton scattering by the
convolution of a handbag diagram and GPD \cite{Rad}.

The confirmation of the importance of FF at NOMAD comes from the
Monte-Carlo simulation reported at this conference \cite{naum}.
The substantial contribution of $\Lambda$s happens to result
from the target remnants even in the current fragmentation region.
Moreover, the qualitative reason for that is the insufficient
energy of $\Lambda$ to break the string, modeling the
fragmentation process \cite{aram}, which corresponds to small $t$
argument discussed above.

As soon as the FF gives the important contribution to cross section,
TOFF should be equally important for T-odd SSA. In this sense,
NOMAD provides the first evidence for TOFF.

As to TOFF role for HERMES, the observed angular
distributions of produced pions do not contain, within the
experimental errors, the term $sin 2\phi$, which is allowed by
the general kinematic analysis, but happens to be compatible
with zero. The expression (\ref{dis}) produces only $sin \phi$,
term, providing the natural explanation of this fact.

In order to compare this approach to the "standard model" of
this effect, which is now probably represented by the
convolution of chiral-odd transversity distribution with chiral-
and T-odd Collins fragmentation function \cite{efr}, one
may try to look for the dependence on the variable $x$ and $z$.
which should be factorizable in that approach at leading order
\cite{leader}.
At the same time, there is no reason for such a factorization in
the case of TOFF. The current level of accuracy, unfortunately,
does not seem to allow for such a check.

\section{Fracture Functions as a framework for distribution and
fragmentation models}

FF (and in particular TOFF) provide a natural framework for understanding
recent suggestions, extending the scope of SIDIS. In particular,
the target spin dependent fragmentation functions,
suggested by M. Gluck and E. Reya \cite{GR}, perfectly fit
to the definition of spin-dependent fracture function.
The criticism  of this paper may therefore be reformulated
as a suggestion about possible role of fracture functions.
Note that because fracture functions include all the information
about the target, the expressions for spin-dependent cross section
should not contain the spin-independent parton distributions anymore.

The model calculation of SSA by Brodsky, Hwang and Schmidt (BHS)\cite{BHS}
may also be related to (T-odd) fracture function. Indeed, their
asymmetry is large only provided the pion transverse momentum is small,
which signals about the possibility of correlation between distribution
and fragmentation functions.

Moreover,
the smallness of transverse momentum
makes standard twist classification unapplicable, as the familiar
twist $3$ suppression factor $M/P_T$ is now not small.
The elegant suggestions of J. Collins \cite{col02}
to attribute BHS asymmetry to gluonic path ordered exponential
may be considered as another manifestation of effective
T-odd distribution \cite{BMT}.
The twist of the effect deserves special discussion.
Although the imaginary phase induced by exponential may appear
already at the leading twist level, it does not change the helicity
structure of the amplitude, and cannot produce the interference and
asymmetry. The latter appears at subleading (according to the standard
counting rules) level and is suppressed as $M/P_T$.
In BHS model this factor, however, is {\it defined} not to be small.

\vspace{0.2cm}
\noindent{\footnotesize
I am most indebted to Organizers for financial support
and warm hospitality.
The work  is supported by grants of RFBR (00-02-16696) and
INTAS (00/587).

\end{document}